\documentclass[twocolumn,showpacs,preprintnumbers,english,amsmath,amssymb]{revtex4}

\usepackage{graphicx,here}% Include figure files
\usepackage{dcolumn}% Align table columns on decimal point
\usepackage{bm}% bold math
\def\simg{\,\hbox{\kern.1em \lower.6ex \hbox{$\sim$} \kern-1.12em
          \raise.6ex \hbox{$>$} }}
\begin{document}

\title{Super-shell structure in harmonically trapped fermionic gases \\ 
       and its semiclassical interpretation}
\date{\today{}}
\author{M. \"{O}gren}
\author{Y. Yu}
\author{S. \AA berg}
\author{S.M. Reimann}
\affiliation{Division of Mathematical Physics, LTH, Lund University. P.O. Box
118, S-221 00 Lund, Sweden}
\author{M. Brack}
\affiliation{Institut f\"{u}r Theoretische Physik, Universit\"{a}t
Regensburg, D-93040 Regensburg, Germany}

\begin{abstract}
It was recently shown in self-consistent Hartree-Fock calculations
that a harmonically trapped 
dilute gas of fermionic atoms with a repulsive two-body interaction 
exhibits a pronounced {\it super-shell} structure: the shell fillings 
due to the spherical harmonic trapping potential are modulated by a 
beat mode. This changes the ``magic numbers'' occurring between the 
beat nodes by half a period. The length and amplitude of the beating mode
depends on the strength of the interaction. We give a qualitative 
interpretation of the beat structure in terms of a semiclassical
trace formula that uniformly describes the symmetry breaking U(3) 
$\rightarrow$ SO(3) in a 3D harmonic oscillator potential perturbed by
an anharmonic term $\propto r^4$ with arbitrary strength. We show that
at low Fermi energies (or particle numbers), the beating gross-shell 
structure of this system is dominated solely by the two-fold degenerate 
circular and (diametrically) pendulating orbits.
\end{abstract}

\maketitle

The realization of Bose-Einstein condensation in trapped 
dilute atom gases \cite{bec}
was a milestone in quantum physics, followed by a  
revolutionary development of both experiment and theory.
Turning from bosonic to fermionic statistics,   
much current interest concerns the 
trapping and cooling of fermionic alkalis. The focus presently often lies
on pairing and the transition to a superfluid state~\cite{becbcs}.

For these ultracold atomic fermi gases, not only is it possible 
to taylor the trap geometry,  but also to 
experimentally change the value of the scattering length 
for two-body collisions~\cite{feshbach}. 
Being able to experimentally modify the interactions
from attractive to repulsive, entirely different interaction regimes can be
probed: Atomic finite fermion systems are a unique laboratory to study 
fundamental quantum phenomena. 

It is apparent that these systems have much in common with atomic nuclei,  
with pronounced shell structures at low particle densities 
and a similar pairing mechanism for {\it attractive} short-range 
interactions~\cite{heiselberg,ben}.
For weak {\it repulsive} interactions between the trapped fermionic atoms, 
shell structure occurs in much analogy also to other 
finite quantal systems, as for example electrons trapped in nanostructured 
semiconductor devices (so-called quantum dots~\cite{steffi}) 
or metallic clusters,  in which the delocalized valence electrons 
are bound in the field of the metallic ions \cite{mbrack}.

Very recently \cite{yongle}, we showed by self-consistent Hartree-Fock 
(HF) calculations that a harmonically trapped gas of fermionic atoms 
interacting by a weak repulsive two-body force may even exhibit 
a so-called super-shell structure. This means that the 
shell oscillations of the spherical harmonic oscillator are 
modulated by a beat structure, whereby the positions of the magic numbers 
are shifted by half a period between successive beats. Similar super-shell 
structure occurs in metallic clusters \cite{cluster}. Inspired by a 
semiclassical analysis of Balian and Bloch in terms of the periodic orbits 
in a spherical cavity \cite{bablo}, supershell beating patterns were 
predicted~\cite{cluster} to occur in the abundance spectra of sodium clusters,
and could later be observed experimentally \cite{klavs}. 

The spherical cavity model of Balian and Bloch~\cite{bablo} 
can, however, not be applied to the present system: 
For a dilute atomic fermi gas with short-range interactions trapped in a
harmonic well, the semiclassical picture is different. 
We show that the super-shell structure originates from  
the interference of diameter 
and circle orbits surviving the breaking of the U(3) symmetry of the 
harmonic oscillator by the leading anharmonicity term in the mean field.

We confine the dilute gas of fer\-mionic atoms by a
spherical harmonic potential modeling an external trap
\cite{grimm}, interacting through a repulsive zero-range two-body
potential $\sim a \sum _{i<j} \delta ^3 (\mathbf{{r}}_{i}-\mathbf{{r}}_{j})$
where $a$ is the $s$-wave scattering length. 
Due to the Pauli principle the $\delta $ interaction only applies to
fermions of pairwise opposite spin. We consider a fully unpolarized
two-component system with two spin states, so that the total particle
density is composed of two different densities of equal magnitude,
$n({\bf r})=n^{\uparrow}({\bf r})+n^{\downarrow}({\bf r})=
2n^{\uparrow}({\bf r})$. In the weak-interaction regime, the
interaction energy density is given by $gn^{\uparrow}({\bf r})
n^{\downarrow}({\bf r})=gn^2({\bf r})\!/4$, where the coupling strength
parameter $g=4\pi\hbar^{2}a/m$ is introduced.
This leads to the single-particle Hartree-Fock equation
\begin{equation}
\left[-\frac{\hbar^{2}}{2m}\Delta+gn^{\uparrow}({\bf r})+V_{ho}({\bf r})\right]
\psi_{i}^{\downarrow}({\bf r})=\epsilon_{i}\psi_{i}^{\downarrow}({\bf r})\,,
\label{HFI}
\end{equation}
where $V_{ho}$ is the harmonic oscillator (HO) trap potential 
(for details, see Ref.~\cite{yongle}). 
In order to treat the interaction as a two-body
process, diluteness of the gas requires that the interparticle 
spacing ${\bar n}^{-1/3}$ is much
larger than the range of the interaction, and that ${\bar n}a^{3}\ll1$.

We solve Eq.\ (\ref{HFI}) self-consistently on a grid under the assumption 
of spherical SO(3) symmetry, which leads to states with $(2l+1)$-fold angular 
momentum degeneracy. The HF interaction term is updated (with some weight 
factors) in each iteration according to $gn^{\uparrow}(r)=
g\sum_i|\psi_{i}^{\uparrow}(r)|^2$.

After convergence is obtained, the HF ground-state energy of the $N$-particle 
system is summed up. In general, the ground-state energy as a function of $N$ 
can be written as the sum of a smooth average part and an oscillating part,
$E_{tot}=E_{av}+E_{osc}$. The oscillating part, referred to as the
shell-correction energy, or shell energy in short, reflects the
quantized level spectrum $\{\epsilon_i\}$. For a non-interacting
Fermi gas in a spherically symmetric $3D$ harmonic trap, the
leading-order term for the average energy is found in the Thomas-Fermi
approximation to be \cite{brack} $E_{av}^{ho}=(3N)^{4/3}\hbar\omega/4$.
For the repulsive interacting case, we find
$E_{av}\left(g>0\right)\propto N^{\alpha}$ with a larger exponent
$\alpha>4/3$. However, Eq.\
(\ref{HFI}) with an interaction term linear in the density is only
valid for moderate $g$ values and in practice we are close to
$\alpha=4/3$ (e.g., $\alpha\approx1.35$ for $g=2$). Contrary to the
non-interacting case, and also to self-saturating fermion systems
(such as nuclei and metal clusters) with a nearly constant particle
density, it is not possible here to obtain the smooth part of the
energy by a simple expansion in volume, surface and higher-order
terms. We therefore perform a numerical averaging of the total HF energy over the particle number $N$ in order to extract its
oscillating part.

\begin{figure}[htbp]
\begin{center}
\includegraphics[angle=0,width=8.5cm]{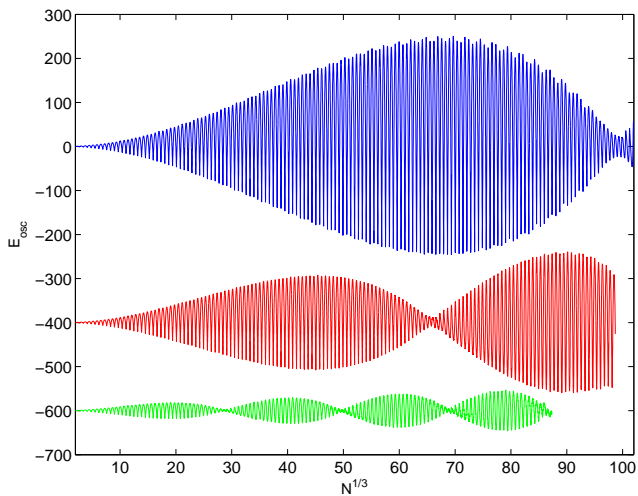}
\end{center}
\vspace*{-0.5cm}
\caption{ The oscillating part of the ground state energy (in units of $\hbar \omega$) as a
function of $N^{1/3}$ for $g= 0.2 $ (top, blue), 0.4 (center, red) and 2 
(bottom, green).
The two lower curves are vertically shifted by $-400\,\hbar\omega$ and $-600
\,\hbar\omega$, respectively. 
%The vertical dotted lines correspond to the HO magic numbers: 
%$N_{mag}^{1/3}=[M(M+1)(M+2)/3]^{1/3}$ for $M=1,2,\dots$
}
\label{figart}
\end{figure}

In the non-interacting case ($g$=0) the shell energy $E_{osc}$
oscillates with a frequency $2\pi\,3^{1/3}\approx 9.06$ as a
function of $N^{1/3}$ and has a smoothly growing amplitude
$\propto N^{2/3}$. This follows from the exact trace formula \cite{brack}
for $E_{osc}$ of the 3D harmonic oscillator, whose leading-order
term is given by
\begin{equation}
E_{osc}^{ho} \simeq (3N)^{\frac23}\frac{\hbar\omega}{2\pi^2}
                    \sum_{k=1}^\infty \frac{(-1)^k}{k^2}\,
                    \cos\left(2\pi k\,(3N)^{\frac13}\right).
\label{tfho}
\end{equation}

Hereby $k$ is the repetition number of the primitive classical periodic
orbit of the system with action $S_0(E)=2\pi E/\omega$. The argument of
the cosine function in Eq.\ (\ref{tfho}) is simply $k$ times $S_0(E)/\hbar$,
taken at the Thomas-Fermi value of the Fermi energy
$E_F(N)=(3N)^{1/3}\hbar\omega$. The gross-shell structure is governed by
the lowest harmonic with $k=1$.

Switching on the interaction, this scenario changes. A beating
modulation of the rapid oscillations is found.
In Fig.\ \ref{figart} we show the shell energy versus $N^{1/3}$ for
three values of the interaction strength, $g$=0.2, 0.4 and 2. A beating
modulation of the amplitude of the shell energy, i.e., a {\it super-shell
structure}, is clearly seen to appear for all cases. At small particle
numbers and particularly for small $g$ values, the shell energy is very 
close to that
of the non-interacting system, given by Eq.\ (\ref{tfho}). For larger
interaction strengths the super-shell structure is more clearly seen,
and several beating nodes appear for $g$=2. With increasing interaction
strength the amplitude of the shell energy oscillations becomes smaller.
For example, for particle numbers around $80^3 \approx 500 000$, the
amplitude of the shell energy is about 40 $\hbar\omega$, which is only
about $10^{-6}$ of the total ground-state energy.

Through Fourier analysis of the calculated shell energy, two frequencies
are seen to smoothly appear with increasing $g$ value around the HO
frequency ($2\pi 3^{1/3}\approx 9.06$).
 The super-shell features appear when the contribution to the
effective potential from the interaction, $g n^{\uparrow}$, is
sufficiently large, i.e., at large values of $g$ and $N$. We also observe
that (almost) until the first super-node, i.e., $N^{1/3}\approx 28 $ in
Fig.\ \ref{figart} (lower curve), the magic numbers agree with the HO ones 
($g=0$).
Between the first two super-nodes, i.e., $ 28\le N^{1/3}\le 49$ in Fig.\
\ref{figart}, the magic numbers for the interacting system are
situated in the middle of two HO magic numbers, i.e., they appear at the
maxima of the fast shell oscillations. Then, after the second super-node
they roughly agree with the unperturbed HO ones again.

In the following we describe the major tools in an ongoing semiclassical 
interpretation of these
features \cite{semicl}. The U(3) symmetry of the unperturbed HO system is
broken by the term $\delta U=gn^\uparrow$ in (\ref{HFI}), resulting in the
SO(3) symmetry of the interacting system. The shortest periodic orbits in
this system are the pendulating diameter orbits and the circular orbits
with a radius corresponding to the minimum of the effective potential 
including the centrifugal term. These two orbits lead to the observed 
supershell beating \cite{semicl}. The above symmetry breaking had not been 
discussed in the semiclassical literature before. In a perturbative
approach \cite{crpert}, it can be accounted for by a group average of the
lowest-order action shift $\Delta S(o)$ brought about by the perturbation
of the system: $\langle e^{\frac{i}{\hbar}\,\Delta S(o)}\rangle_{o\in U(3)}$.
Hereby $o$ is an element of the group U(3) characterizing a member of the
unperturbed HO orbit family (ellipses or circles). For the average it is
sufficient to integrate over the 4-dimensional manifold $\mathbb{C}$P$^2$ 
\cite{bbz}, which for a perturbation $\delta U(r) = \varepsilon r^4$ can be 
done analytically \cite{semicl}. We therefore model the self-consistent
numerical HF field by the following perturbed HO potential:
\begin{equation}
V(r) = V_{ho}(r) + \frac{\epsilon}{4}\,r^4\,,
\label{modpot}
\end{equation}
where the anharmonic term simulates the symmetry breaking effect of the part $gn^{\uparrow}
({\bf r})$ in (\ref{HFI}). For small interaction strenghts $g$, $\epsilon$ 
is proportional to $g$.

In the perturbative regime ($\varepsilon\ll 1$) we have found the following
perturbed trace formula for the level density:
\begin{eqnarray}
\delta g_{pert}(E) = \frac{\omega^2}{2\varepsilon\hbar^2\pi}\!
                    \sum_{k=1}^\infty \frac{(-1)^k}{k}\!
                    \left[\sin\!\left(\!\frac{kS_c}{\hbar}\!\right)\!
                    -\sin\!\left(\!\frac{kS_d}{\hbar}\!\right)\!\right]\!\!,
\label{tfpert}
\end{eqnarray}
where $kS_d$ and $kS_c$ are the classical actions of the diameter and
circle orbits, respectively. In the limit $\varepsilon\rightarrow0$,
their difference goes as $k(S_c-S_d)\rightarrow k\varepsilon\pi
E_F^2(N)/\omega^5$, so that (\ref{tfpert}) tends to the level density of 
the pure HO limit corresponding to (\ref{tfho}). 

To cover larger values of $g$ and $N$, we have also developed an analytical 
uniform trace formula \cite{semicl}. Uniform here means that it reproduces
the HO trace formula for the U(3) (higher symmetry) limit $\epsilon=0$ in 
a smooth way, analogously to those derived earlier for U(1) \cite{toms} and 
U(2) symmetry breaking \cite{hhuni}. In the large-energy (or large-$N$) 
limit, it yields the correct trace formula for the diameter and circular 
orbits, forming two-fold degenerate families, valid for all strenghts
$\epsilon$:
\begin{equation}
\delta g(e) \simeq \sum_{k=1}^\infty \left[
            {\cal A}_k^d(e)\sin(kS_d(e)/\hbar)+
            {\cal A}_k^c(e)\sin(kS_c(e)/\hbar)\right]\!.
\label{dguni}
\end{equation}
Analytical expressions for the amplitudes and actions (in terms of elliptic 
integrals) are given in \cite{semicl}. Eq.\ (\ref{dguni}) goes over into 
the perturbative trace formula (\ref{tfpert}) in the limit 
$\epsilon\rightarrow 0$. This confirmes the statement made in \cite{yongle} 
that only the diameter and circle orbits are important in this limit.

\begin{widetext}

\begin{figure}[H]
\begin{center}\includegraphics[%
  width=14.5cm,
  angle=0]{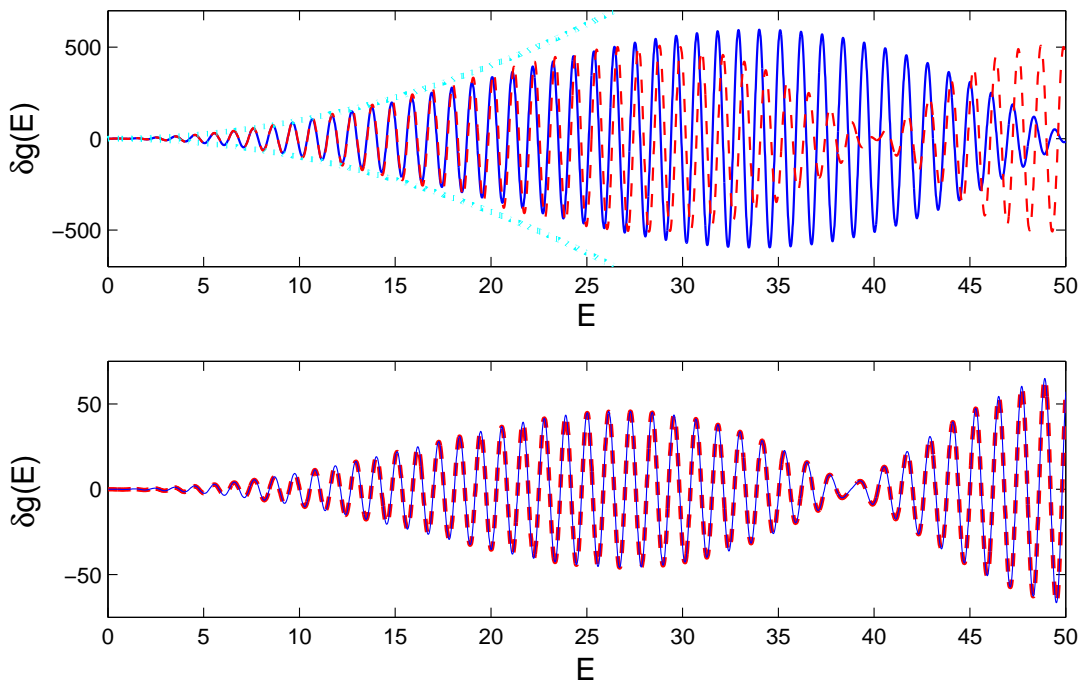}\end{center} %7.9 dgpert.eps
\vspace*{-0.7cm}
\caption{Oscillating part $\delta g(E)$ of level density of the perturbed 
HO potential (\ref{modpot}) versus energy $E$ (unit: $\hbar\omega$), 
coarse-grained by Gaussian convolution over an energy range $\gamma=0.5
\hbar\omega$. 
{\it Upper panel:} comparison of perturbative trace formula (\ref{tfpert}) 
(dashed line, red) and uniform trace formula (\ref{dguni}) (solid line, 
blue) for the value $\epsilon=0.005$. The dotted lines describe the 
amplitude for the unperturbed HO limit $\epsilon=0$. 
{\it Lower panel:} comparison of uniform trace formula (\ref{dguni}) 
(solid line, blue) and exact quantum-mechanical results (dashed line, red)
for $\epsilon=0.01$.}
\label{figart1g1}
\end{figure}

\end{widetext}

In the upper panel of Fig.\ \ref{figart1g1}, we compare the 
%coarse-grained oscillating parts
%of the 
 level density for the potential (\ref{modpot}) obtained with the
perturbative trace formula (\ref{tfpert}) (dashed line, red) and the
uniform trace fomula (\ref{dguni}) (solid line, blue) for the value  
$\epsilon=0.005$, using only the lowest harmonics ($k=1$). While
there is good agreement up to the first super-shell maximum for this
small value of $\epsilon$, for larger perturbations the two formulas 
only agree for very small energies. In the lower panel of Fig.\ 
\ref{figart1g1} we compare the coarse-grained result of the uniform trace formula 
(\ref{dguni}) (solid line, blue) with the exact quantum-mechanical result 
(dashed line, red) for $\epsilon=0.01$, demonstrating
that the semiclassical interpretation of the super-shell beat in terms
of circular and diameter orbits is perfect.

At sufficiently high energies and perturbation strengths $\epsilon$, 
three-fold degenerate families of tori with rational ratios 
$\omega_r:\omega_\varphi=n:m\geq 7:3$ of radial and angular frequency 
bifurcate from higher repetitions ($k\geq 3$) of the circle orbit 
\cite{arita}. The shortest of them, a star-like orbit with $n:m=7:3$, 
is approximately 10 times longer than the shortest diameter and circle 
orbits. Therefore, all the tori only contribute to finer quantum 
structures at higher energies. They can be included in the semiclasscial 
trace formula using standard techniques \cite{bertab,crli,kaidel}.

The beat structure in $E_{osc}$ has some similarities with that found 
in nuclei \cite{bm} and metal clusters \cite{mbrack}. There are, however, 
two essential differences. 1. Those systems are self-saturating and have 
steep mean-field potentials that can be modeled by a spherical cavity 
\cite{bablo}. The present system, in contrast, has a mean field with much 
smoother walls that are dominated at large distances by the confining 
harmonic potential. 2. The super-shells in the cavity model come from the 
interference of the shortest periodic orbit families with three-fold
degeneracy, as is usual in spherical systems \cite{bertab,crli}. Here, 
however, the gross-shell structure comes uniquely from the diameter and 
circle orbits which are only two-fold degenerate, whereas the fully
three-fold degenerate tori are so much longer that they only affect
the finer quantum structures at higher energies.

In conclusion, we have seen that the shell structure of weakly
interacting fermions in a harmonic trap shows a pronounced beating
pattern, with the single shell positions changing by half a period
length between the different beat nodes. A Fourier analysis of the
oscillating shell-correction part of the Hartree-Fock energy shows
clear peaks at two slightly different frequencies. This we have 
interpreted semiclassically by the interference of the shortest
periodic orbits generated by the breaking of the U(3) symmetry of
the non-interacting HO system, which are the families of diameter
and circle orbits. A more detailed quantitative interpretation using 
the uniform trace formula (\ref{dguni}) derived in \cite{semicl} is 
in progress. After extracting $\varepsilon$ and $\omega_{eff}$ from a 
polynomial fit to the numerical HF potential, we expect to describe 
the beat structure in the numerically obtained HF shell energies 
$E_{osc}(N)$ quantitatively in terms of classical periodic orbits.

This work was partially financed by the Swedish Foundation for 
Strategic Research (SSF) and the Swedish Research Council (VR). 
M.\,\"{O}.\ acknowledges financial support from the Deutsche 
Forschungsgesellschaft (Graduiertenkolleg 638: {\it Nichtlinearit\"at und
Nichtgleichgewicht in kondensierter Materie}) and the warm hospitality 
at the Universit\"{a}t Regensburg during several research visits.


\begin{thebibliography}{99}

\bibitem{bec} K.B. Davis, M.O. Mewes, M.R. Andrews, N.J. von Druten,
  D.S. Durfee, D.M. Kurn and W. Ketterle, 
Phys. Rev. Lett. {\bf 75}, 3969 (1995); 
M.H. Anderson, J.R. Ensher, M.R. Matthews, C.E. Wieman and
  E.A. Cornell, Science {\bf 269}, 198 (1995).

\bibitem{becbcs} C. A. Regal, M. Greiner, and D. S. Jin,
Phys. Rev. Lett. {\bf 92}, 040403 (2004);
M. W. Zwierlein {\it et al.}, Phys. Rev. Lett. {\bf 92}, 120403 (2004);
J. Kinast {\it et al.}, Phys. Rev. Lett. {\bf 92}, 150402 (2004);
Science {\bf 307}, 1296 (2005); M. Bartenstein {\it et al.},
Phys. Rev. Lett. {\bf 92}, 120401 (2004), Phys. Rev. Lett. {\bf 92}, 
203201 (2004); T. Bournel {\it et al.}, Phys. Rev. Lett. {\bf 93}, 
050401 (2004). 

\bibitem{heiselberg} H. Heiselberg, preprint arXiv:cond-mat/0304005 (2003).

\bibitem{ben} H. Heiselberg and B. Mottelson, Phys. Rev. Lett. {\bf 88},
  190401 (2002).

\bibitem{feshbach} E. Timmermans, P. Tommasini, M. Hussein and A. Ker\-man,
                   Phys.\ Rep.\ {\bf 315}, 199 (1999) and refs.\ therein.

\bibitem{bm} Aa. Bohr and B. R. Mottelson: {\it Nuclear Structure}
             Vol. I (World Scientific, New York, 1975).

\bibitem{mbrack} M. Brack, Rev.\ Mod.\ Phys.\ {\bf 65}, 677 (1993);
                 The Scientific American, December 1997, p.\ 50.

\bibitem{steffi} S. M. Reimann and M. Manninen,
                 Rev.\ Mod.\ Phys.\ {\bf 74}, 1283 (2002).

\bibitem{grimm} R. Grimm, M. Weiderm\"{u}ller and Y. B. Ovchinnikov,
                page 95-170 in
                {\it Advances in Atomic, Molecular and Optical Physics}
                (Academic Press, 2000).

\bibitem{yongle} Y. Yu, M. \"Ogren, S. \AA berg, S. M. Reimann and M. Brack,
                   preprint arXiv:cond-mat/0502096 (2005).

\bibitem{cluster} H. Nishioka, K. Hansen, and B. R. Mottelson,
                  Phys.\ Rev.\ B {\bf 42}, 9377 (1990).

\bibitem{bablo} R. Balian and C. Bloch, Ann.\ Phys.\ (N.Y.)
                {\bf 69}, 76 (1972).

\bibitem{klavs} J. Pedersen, S. Bj{\o}rnholm, J. Borggren, K. Hansen,
                T. P. Martin, and H. D. Rasmussen, Nature {\bf 353},
                733 (1991).

\bibitem{wire} A. I. Yanson, I.K. Yanson, and J.M. van Ruitenbeek,
               Nature {\bf 400}, 144 (1999); Phys.\ Rev.\ Lett.\
               {\bf 84}, 5832 (2000).

\bibitem{petrov} D. S. Petrov, Phys.\ Rev.\ A \textbf{67}, 010703
                 (2003).

\bibitem{brack} M. Brack and R. K. Bhaduri: {\it Semiclassical
                Physics} (Westview Press, Boulder, USA, 2003).

\bibitem{semicl} M. Brack, M. \"Ogren, Y. Yu and S. M. Reimann, 
                 preprint arXiv:nlin.SI/0505060 (2005).

\bibitem{crpert} S. C. Creagh, Ann.\ Phys.\ (N. Y.) {\bf 248}, 60 (1996).

\bibitem{bbz}  see, e.g., I. Bengtsson, J. Br\"annlund, and K. \.{Z}yczkowski,
               Int.\ J. Mod.\ Phys.\ A {\bf 17}, 4675 (2002).

\bibitem{toms} S. Tomsovic, M. Grinberg, and D. Ullmo, Phys.\ Rev.\ Lett.\
                {\bf 75}, 4346 (1995).

\bibitem{hhuni} M. Brack, P. Meier, and K. Tanaka, J. Phys.\ A {\bf 32}, 331
                (1999).

\bibitem{arita} This situation persists in the limit $\epsilon\to\infty$, 
                which corresponds to a pure quartic oscillator potential,
                and is consistent with a general result obtained for
                homogeneous potentials $V(r)=cr^\alpha$ with $\alpha\geq 2$,
                see K. Arita, Int.\ J.\ Mod. Phys. E {\bf 13}, 191 (2004).

\bibitem{bertab}  M. V. Berry and M. Tabor, Proc.\ R. Soc.\ Lond.\ A {\bf 349},
                  101 (1976); J. Phys.\ A {\bf 10}, 371 (1977).

\bibitem{crli} S. C. Creagh and R. G. Littlejohn, Phys.\ Rev.\ A {\bf 44}, 836
               (1991); J. Phys.\ A {\bf 25}, 1643 (1992).

\bibitem{kaidel} J. Kaidel, M. Brack, Phys.\ Rev.\ E {\bf 70}, 016206 (2004).

\end{thebibliography}
\end{document}